# Energy Control of Grid-forming Energy Storage based on Bandwidth Separation Principle

Chu Sun, *Member, IEEE*, Syed Qaseem Ali, *Member, IEEE*, Geza Joos, *Life Fellow, IEEE*

*Abstract*- The reduced inertia in power system introduces more operation risks and challenges to frequency regulation. The existing virtual inertia and frequency support control are restricted by the normally non-dispatchable energy resources behind the power electronic converters. In this letter, an improved virtual synchronous machine (VSM) control based on energy storage is proposed, considering the limitation of state-of-charge. The steady-state energy consumed by energy storage in inertia, damping and frequency services is investigated. Based on bandwidth separation principle, an energy recovery control is designed to restore the energy consumed, thereby ensuring constant energy reserve. Effectiveness of the proposed control and design is verified by comprehensive simulation results.

*Keywords*- Virtual synchronous machine, energy storage, energy consumption, energy recovery, bandwidth separation.

## I. INTRODUCTION

With rechargeable and partially controllable capability and high response speed, energy storage system (ESS) is a promising candidate for implementing VSM or grid-forming converter in power system [1]. However, the state-of-charge (SoC) of ESS should be retained within an acceptable range for secure and sustainable operation. A SoC feedback control for VSM based on hybrid storage is proposed in [2] to avoid over-charging/discharging, but such feedback is SoC limit triggered, which cannot ensure constant reserve for continuous operation. The authors of [3] put forth an improved VSM control based on type-IV wind turbine generator and short-term ESS whose SoC is managed by additional proportional-integral (PI) control. However, the energy variation of ESS, and the coordination between SoC management and frequency regulation are not discussed.

Based on the literature review and our previous work [4], a general scheme of fitting the SoC energy recovery control of ESS into power system frequency control is introduced in this letter. The steady-state energy variation of VSM in different services are quantified, and bandwidth separation principle is adopted in frequency and SoC control loop design.

## II. VSM WITH SOC RECOVERY CONTROL

A general power network is shown in Fig. 1, consisting of a synchronous generator (SG), an energy storage system, and connected networks with other normally non-dispatchable devices such as renewable energy generators and loads. Compared with conventional VSM, a SoC recovery control based on PI regulator is added in the power reference of ESS. The virtual inertia, damping and governor control of VSM can be assigned to hybrid storage systems under DC or AC coupled scheme or a single ESS (normally fasting-acting ESS) [4]. The different implementation of VSM control is equivalent in terms of frequency regulation, and represented uniformly in Fig. 2.

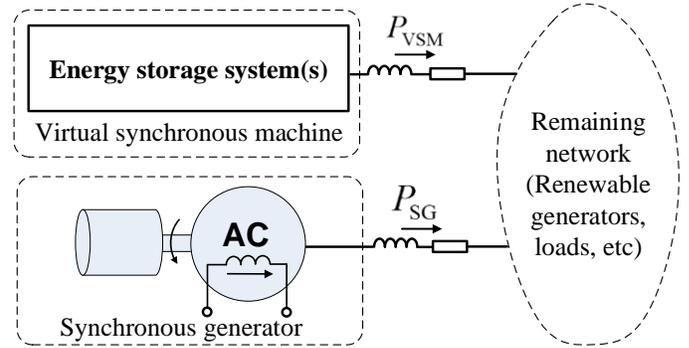

Fig. 1. Power system with synchronous generator and ESS-based VSM.

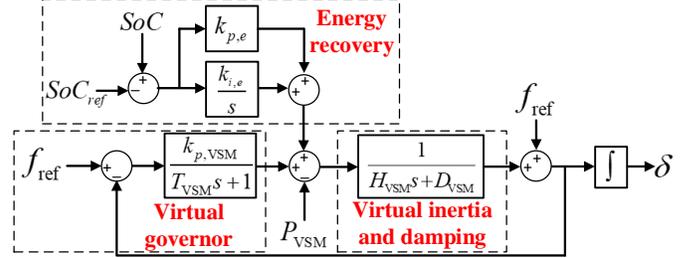

Fig. 2. Diagram of ESS-based VSM with energy recovery control.

## III. DYNAMIC ANALYSIS AND CONTROL DESIGN

### A. Modeling of Power, Frequency and Energy Dynamics

The power-frequency dynamics due to the load power disturbance $\Delta P_L$ can be estimated by (1)-(4), in which $H$ and $D$ represent the inertia and damping, $k_p$ and $k_i$ are the governor control gain and integral gain of frequency control, the subscripts "SG" and "VSM" correspond to synchronous generator and virtual synchronous machine, respectively. $\Delta P_{SG}$, $\Delta P_{HD}$, $\Delta P_{Gov}$ represent the power response of SG, virtual inertia and damping, and virtual governor, respectively.

$$\Delta f(s) = G_f(s)\Delta P_L =$$

$$-\frac{(T_{SG}s+1)(T_{VSM}s+1)s}{\begin{pmatrix}((H_{SG}+H_{VSM})s+D_{SG}+D_{VSM})(T_{SG}s+1)(T_{VSM}s+1)s \\ +k_{p,VSM}s(T_{SG}s+1)+(k_{p,SG}s+k_{i,SG})(T_{VSM}s+1)\end{pmatrix}}\Delta P_L \quad (1)$$

$$\Delta P_{SG}(s) = -\left(\frac{k_{p,SG}+k_{i,SG}/s}{T_{SG}s+1} + (H_{SG}s+D_{SG})\right)\Delta f(s) \quad (2)$$

$$\Delta P_{HD}(s) = -(H_{VSM}s+D_{VSM})\Delta f(s) \quad (3)$$

$$\Delta P_{\text{Gov}}(s) = -k_{p,\text{VSM}}\Delta f(s)/(T_{\text{VSM}}s+1) \quad (4)$$

By final-value theorem, the steady-state energy variation of ESS in virtual inertia and damping ($\Delta E_{\text{HD}}$), and governor control ($\Delta E_{\text{Gov}}$), is estimated by (5). If all the control is implemented on one ESS, the total steady-state energy variation of VSM is given by (6). The energy variation can be used as criteria for sizing ESS in VSM applications [4].

$$\Delta E_{\text{HD}} = \lim_{s\to 0} s\Delta E_{\text{HD}}(s) = \Delta P_{\text{HD}}(s)|_{s\to 0} = (D_{\text{VSM}}/k_{i,\text{SG}})\Delta P_L \quad (5)$$
$$\Delta E_{\text{Gov}} = \lim_{s\to 0} s\Delta E_{\text{Gov}}(s) = \Delta P_{\text{Gov}}(s)|_{s\to 0} = (k_{p,\text{VSM}}/k_{i,\text{SG}})\Delta P_L$$

$$\Delta E_{\text{VSM}} = (D_{\text{VSM}} + k_{p,\text{VSM}})\Delta P_L / k_{i,\text{SG}} \quad (6)$$

### B. Bandwidth Separation in Frequency and SoC Control

Letting $s=0$ in governor control, the simplified model of the primary frequency control is shown in Fig. 3(a) where the total inertia and damping of the system are denoted by $H_t=H_{\text{SG}}+H_{\text{VSM}}$ and $D_t=D_{\text{SG}}+D_{\text{VSM}}$, respectively. Therefore, the bandwidth of primary control ($\omega_{\text{bw,primary}}$) is approximated by

$$\omega_{\text{bw,primary}} = (k_{p,\text{SG}} + k_{p,\text{VSM}} + D_{\text{SG}} + D_{\text{VSM}})/(H_{\text{SG}} + H_{\text{VSM}}) \quad (7)$$

For secondary frequency control, its bandwidth ($\omega_{\text{bw,secondary}}$) can be evaluated by $G_f(s)/s$ [5]. Since $\omega_{\text{bw,secondary}}$ should be much lower than the primary control, its model can be simplified as Fig. 3(b). $\omega_{\text{bw,secondary}}$ is then estimated as

$$\omega_{\text{bw,secondary}} = k_{i,\text{SG}} / (k_{p,\text{SG}} + k_{p,\text{VSM}} + D_{\text{SG}} + D_{\text{VSM}}) \quad (8)$$

The overall frequency control considering storage management is displayed in Fig. 3(c), where $k_{p,e}$ and $k_{i,e}$ are the proportional and integral gain of SoC recovery control, $E_{\text{nom}}$ and $SoC_{\text{ini}}$ are the energy capacity and the initial SoC of ESS, respectively. The SoC recovery control loop can be transformed into Fig. 4 where the effect of frequency control appears in the disturbance. Since the control plant is an integral block, a proportional block should normally be able to achieve zero steady-state error. As the outermost loop, the bandwidth ($\omega_{\text{bw,SoC}}$) of the simplified SoC control loop $G_e(s)$ is then given by (9), which should be lower than that of frequency regulation.

$$\omega_{\text{bw,SoC}} = k_{p,e} / E_{\text{nom}} \quad (9)$$

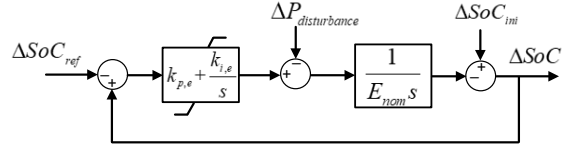

Fig. 4. Diagram of the simplified SoC recovery control for ESS.

A larger value of $k_{p,e}$ can achieve faster energy recovery, but may undermine frequency performance. Considering the parasitic losses of ESS, an integral block is added in the SoC recovery control. To ensure proper damping and keep the original bandwidth, a small $k_{i,e}$ (0.002) is selected here.

For the parameters in Table I, the estimated bandwidth of the three control loops are shown in Fig. 5, which clearly obeys the bandwidth separation rule. Those values obtained from the Bode plots of $G_e(s)$ and $G_f(s)$ also closely matches the analytical estimation, as illustrated in Fig. 6. Notice that the bandwidth of energy recovery control should be higher than that of third frequency control ($\omega_{\text{bw,third}}$) to avoid the interference into steady-state energy management.

TABLE I. PARAMETERS AND VALUES OF THE TESTED SYSTEM

| Parameters | Values | Parameters | Values |
|---|---|---|---|
| Base Voltage $f_{ref}$ | 60 Hz | Base power | 320 kVA |
| Base voltage $V_{ac\,(\text{LL-rms})}$ | 600 V | $E_{\text{nom}}$ | 6.8 p.u.·s |
| Power rating of ESS | 1 p.u. | $SoC_{ref}$ of ESS | 0.5 |
| $T_{\text{VSM}}$ | 0.3 s | $H_{\text{VSM}}$ | 5 s |
| $k_{p,\text{VSM}}$ | 15 p.u. | $D_{\text{VSM}}$ | 10 p.u. |
| $k_{p,e}$ | 0.4 p.u. | $k_{i,e}$ | 0.002 |
| Power rating of SG | 1 p.u. | $H_{\text{SG}}$ | 2.5 s |
| $D_{\text{SG}}$ | 0 | $T_{\text{SG}}$ | 0.3 s |
| $k_{p,\text{SG}}$ | 15 p.u. | $k_{i,\text{SG}}$ | 5 p.u. |

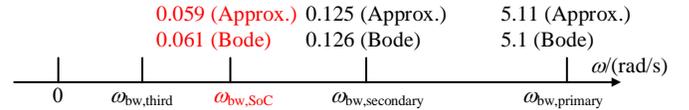

Fig. 5. Illustration of the bandwidth of frequency and SoC control loops.

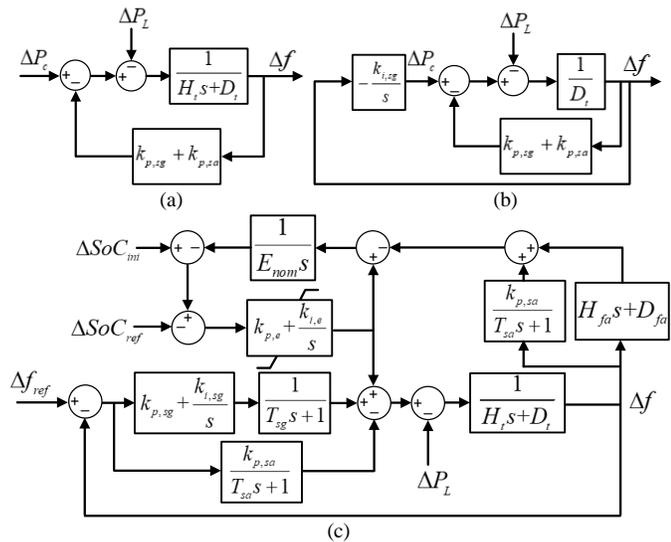

Fig. 3. Diagram of simplified frequency control. (a) Simplified primary control. (b) Simplified secondary control. (c) Integrated with SoC recovery control.

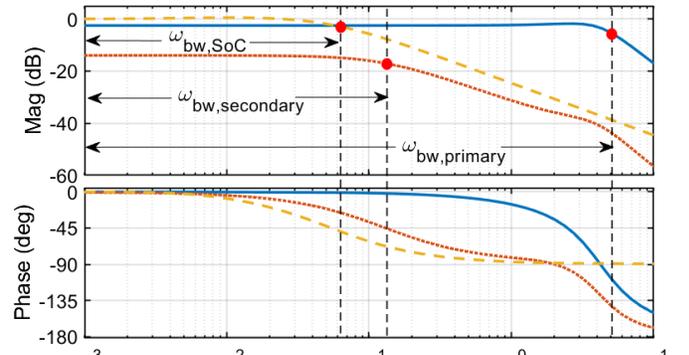

Fig. 6. Bode plots illustrating the bandwidth of primary frequency, secondary frequency and SoC recovery control.

## IV. VERIFICATION OF VSM CONTROL

A system with one SG and one ESS-based VSM serving active loads as in Fig. 1 is simulated. First, no VSM is present in the system and the SG supplies the total loads. For a 0.375 p.u. load increase at $t=10$ s, the frequency nadir is 57.87 Hz at $t=10.47$ s, as Fig. 7 shows. The large frequency deviation results from the small inertia of SG. The swing in the frequency waveform is relatively severe due to the low damping provided by SG.

When VSM control without energy recovery is applied to ESS, the frequency nadir is 59.35 Hz at $t$=10.39 s for the same load change, as shown in Fig. 8. The transient power released from ESS in virtual inertia, damping and governor control follows the high-pass-filter (HPF) and low-power-filter (LPF) processes, in consistent with the transfer functions in (3) and (4). The oscillation in frequency is mitigated due to the damping from ESS. However, the SoC of ESS keeps dropping, as with conventional VSM or grid-forming inverter. The SoC variation (24.63%) is close to the analytical estimation (27.57%) based on (6). Such deep discharging will intensify ESS aging for long-time operation.

After adding SoC recovery control to the ESS, the frequency nadir becomes 59.34 Hz at $t$=10.38 s, as displayed in Fig. 9, which approximates the results in Fig. 8. The power of ESS gradually decreases to zero and its SoC returns to 0.5 after frequency control. For the selected parameters in the SoC recovery control, the impact on frequency waveform is almost negligible. The settling time for SoC recovery is around 100 s, which is close to the bandwidth estimation in (9).

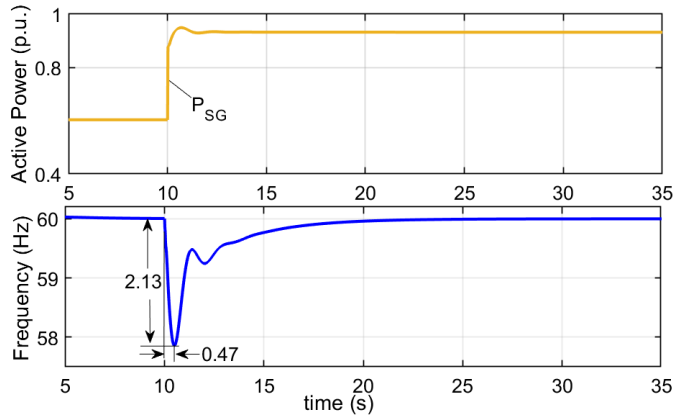

Fig. 7. Results for synchronous generator solely serving the loads.

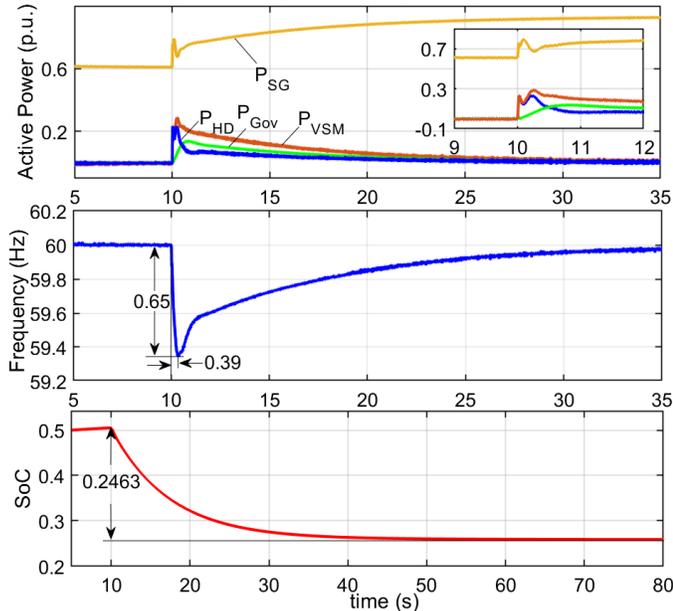

Fig. 8. Test results for the case of adding FAESS without SoC recovery control. (a) Active power. (b) Frequency. (c) SoC of the ESS.

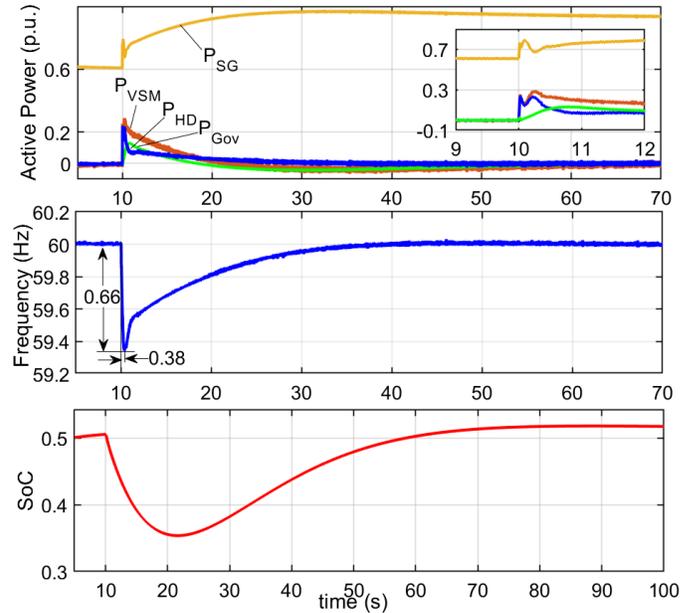

Fig. 9. Test results for the case of adding FAESS with SoC recovery control. (a) Active power. (b) Frequency. (c) SoC of the ESS.

## V. CONCLUSION

This paper introduced a general framework of involving the VSM-controlled ESS with small energy capacity in frequency regulation, in the context of grid-forming converters in power system. The steady-state energy variation of ESS is proportional to the virtual damping and governor control gain, which demonstrates the need of energy recovery control. The energy recovery control designed based on bandwidth separation can restore the energy of ESS, thus reducing the size and cost of ESS without undermining frequency control performance. The VSM with energy recovery control can be termed as "virtual synchronous condenser", given energy-neutral features in the inertial and damping services.

## REFERENCE

[1] Q. Zhong, Y. Wang and B. Ren, "Connecting the Home Grid to the Public Grid: Field Demonstration of Virtual Synchronous Machines," in IEEE Power Electronics Magazine, vol. 6, no. 4, pp. 41-49, Dec. 2019..

[2] Y. Ma, W. Cao, L. Yang, F. Wang and L. M. Tolbert, "Virtual Synchronous Generator Control of Full Converter Wind Turbines with Short-Term Energy Storage," in IEEE Transactions on Industrial Electronics, vol. 64, no. 11, pp. 8821-8831, Nov. 2017.

[3] M. Shi, H. Chen, C. Zhang, F. Mei, J. Fang and H. Miao, "A Virtual Synchronous Generator System Control Method with Battery SOC Feedback," 2018 2nd IEEE Conference on Energy Internet and Energy System Integration (EI2), Beijing, 2018.

[4] C. Sun, S. Q. Ali, G. Joos and F. Bouffard, "Design of Hybrid-Storage-Based Virtual Synchronous Machine With Energy Recovery Control Considering Energy Consumed in Inertial and Damping Support," in IEEE Transactions on Power Electronics, vol. 37, no. 3, pp. 2648-2666, March 2022.

[5] Fan, L. (2017). Control and Dynamics in Power Systems and Microgrids (1st ed.). CRC Press.